\documentclass[12pt]{iopart}

\usepackage{iopams}  
\usepackage{cite}
\usepackage{amsthm}
\usepackage{hyperref}
\usepackage{graphicx}
\usepackage{float}
\usepackage{csquotes}
\usepackage{nowidow}

\newtheorem{theorem}{Theorem}
\newtheorem*{remark}{Remark}

\graphicspath{{./plots/}{./sketch/}}

\newcommand{\C}{\mathbb{C}}  
\newcommand{\R}{\mathbb{R}}  
\newcommand{\D}{D}  
\newcommand{\Dbar}{\overline{D}}
\newcommand{\Omegabar}{\overline{\Omega}}
\newcommand{\Sone}{S^1} 
\newcommand{\Jac}{\mathcal{J}}  
\newcommand{\solspace}{C\left(\overline{\Omega}\right) \cap C^2\left(\Omega\right)}  
\newcommand{\Id}{\mathrm{Id}}  
\newcommand{\xy}{\left(x,y\right)}  
\newcommand{\xieta}{\xi+\rmi\eta}  
\newcommand{\rhotheta}{\rho\rme^{\rmi\theta}}  

\begin{document}
\title[Boundary conforming coordinates for arbitrarily shaped toroidal domains.]{Construction of an invertible mapping to boundary conforming coordinates for arbitrarily shaped toroidal domains.}
\author{Robert Babin$^{1,2}$, Florian Hindenlang$^{1}$,  Omar Maj$^{1}$ and Robert Köberl$^{1,2}$}
\address{$^1$ Numerical Methods in Plasma Physics, Max Planck Institute for Plasma Physics, Garching, Germany}
\address{$^2$ School of Computation, Information and Technology, Technical University of Munich, Munich, Germany}
\ead{robert.babin@ipp.mpg.de}


\begin{abstract}
    Boundary conforming coordinates are commonly used in plasma physics to describe the geometry of toroidal domains, for example, in three-dimensional magnetohydrodynamic equilibrium solvers.
    The magnetohydrodynamic equilibrium configuration can be approximated with an inverse map, defining nested surfaces of constant magnetic flux.
    For equilibrium solvers that solve for this inverse map iteratively, the initial guess for the inverse map must be well defined and invertible.
    Even if magnetic islands are to be included in the representation, boundary conforming coordinates can still be useful, for example to parametrize the interface surfaces in multi-region, relaxed magnetohydrodynamics
    or as a general-purpose, field-agnostic coordinate system in strongly shaped domains.
    Given a fixed boundary shape, finding a valid boundary conforming mapping can be challenging, especially for the non-convex boundaries from recent developments in stellarator optimization.
    In this work, we propose a new algorithm to construct such a mapping, by solving two Dirichlet-Laplace problems via a boundary integral method. We can prove that the generated harmonic map is always smooth and has a smooth inverse. 
    Furthermore, we can find a discrete approximation of the mapping that preserves this property.
\end{abstract}
\noindent{\it Keywords\/}: boundary conforming coordinates, diffeomorphism, harmonic map, MHD, equilibrium

\submitto{\PPCF}
\maketitle


\section{Introduction} \label{sec:intro}
In plasma physics for magnetic confinement fusion, the geometry of the plasma is often described by a toroidal domain, diffeomorphic to the solid torus.
That is, there exists a differentiable and invertible mapping from the toroidal domain to the solid torus, with its inverse being differentiable as well.
Besides cylindrical coordinates, other descriptions of the toroidal domain, like the Frenet-Serret frame or flux-aligned coordinates, are used.
The latter are a type of boundary conforming coordinates, where one coordinate is radius-like and increases from the magnetic axis to the boundary of the domain, and the other two coordinates are angle-like and periodic with period $2\pi$.
In particular, for flux-aligned coordinates, the radial coordinate is constant on a family of nested toroidal surfaces on which the magnetic field is tangent, referred to as flux surfaces, and the angle-like coordinates parameterize these flux surfaces.
Such coordinates are especially suited for spectral methods, e.g. employing Fourier-polynomial decomposition.
Several state-of-the-art 3D magnetohydrodynamic (MHD) equilibrium solvers like VMEC\cite{hirshman_VMEC_1983}, DESC\cite{dudt_DESC_2020} or GVEC\cite{hindenlang_GVEC_2019,hindenlang_Frenet_2024} describe the magnetic field and flux surface geometry (in absence of magnetic islands) using an inverse coordinate representation, i.e. by describing the flux surface positions in real space (e.g. in cartesian or cylindrical coordinates) as functions of the flux-aligned coordinates.
Such codes find the equilibrium solution by iteratively changing the coordinate map, e.g. in order to minimize energy.
Therefore it is crucial that the initial mapping from flux-aligned coordinates to real space is well defined.
The multi-region relaxed MHD equilibrium solver SPEC\cite{Hudson_SPEC_2012}, which can include magnetic islands in the equilibrium, does not use flux-aligned coordinates, but still relies on a discrete number of interface surfaces that need to be nested.

Recent developments in stellarator optimization using near-axis expansion\cite{garren_Magnetic_1991, garren_Existence_1991, landreman_mapping_2022} have resulted in strongly shaped plasma configurations.
It is now of interest to explore configurations with similarly shaped plasma boundaries using higher-fidelity 3D MHD equilibrium solvers.
However, initialization of equilibrium codes can be challenging with such boundaries, as conventional radial blending techniques (see \ref{app:radial_blending}) between the plasma boundary and a specified initial magnetic axis fail.
That is, they lead to overlapping flux surfaces and, thus, an invalid coordinate mapping.
In addition, the position of the magnetic axis is a result of the equilibrium calculation and, as such, should not be needed for the initialization.
There are algorithms to find an initial magnetic axis\cite{qu_Coordinate_2020}, but there are configurations where radial blending fails with any chosen axis.
From a mathematical point of view, we are interested in finding boundary conforming coordinates for arbitrarily shaped toroidal domains.

As we only consider toroidal domains, we can further reduce the problem by one dimension and find two-dimensional boundary conforming coordinates for each cross-section of the toroidal domain.
The domain is then composed of cross-sections that are parametrized by the toroidal angle.
Finding boundary conforming coordinates for each cross-section is equivalent to finding a diffeomorphism from a two-dimensional, simply connected domain to the unit disk.

Recently a method was proposed to find a mapping from the solid torus to the considered domain using a variational principle which minimizes the squared Jacobian determinant of the domain with a penalty on the curvature of radial coordinate lines \cite{tecchiolli_constructing_2024}.

A similar problem is found in the context of boundary-fitted mesh generation, where a valid and smooth mesh needs to be generated in a domain that is only defined by its boundary. The boundary can be curved and also non-convex. A successful approach is the so-called elliptic mesh generator \cite{thompson82}, which solves elliptic PDEs in order to generate the mesh. In particular, a well-known tool for structured grids is the solution of the Winslow equations, which results in so-called Laplace or Harmonic grids, first introduced by Winslow\cite{winslow1966numerical}, and studied extensively since then, for example in \cite{knupp92,thompson82}. A variational formulation for the Winslow equations exists\cite{winslow_variational} and has been applied to unstructured high-order mesh generation\cite{fortunato_persson}. The Winslow equations consist of two Laplace equations defined on the physical domain, and solving for the logical coordinates in order to find the mapping to the logical domain. Often, the Winslow equations are transformed to logical space, resulting in a nonlinear PDE for the mapping, and therefore, must be solved iteratively.

A simple solution that is guaranteed to exist for such a problem is a conformal map, for which there are also construction methods available.
The conformal map is angle preserving and orthogonal, which is a strong constraint on the mapping and is often ill-suited for discretization, as can be seen for a simple domain bounded by an ellipse, discussed in \sref{sec:convergence}.
In contrast, the elliptic domain can also be expressed by a linear transformation of the unit disk, discussed in \sref{sec:simple-example-ellipse}, which is not a conformal map, but easily discretized.
Both the linear map and the conformal map are special cases of so called harmonic maps.
In the case of planar domains, a harmonic mapping is defined by two coordinate functions that are harmonic\cite{duren_Harmonic_2004}.
That is, they are twice differentiable functions that satisfy Laplace's equation in the whole domain.
While the conformal map is guaranteed to be invertible, a more general harmonic map is only guaranteed to be invertible if one maps the unit circle to a convex domain\cite{kneser_A41_1926}.
If one maps the unit circle to a non-convex domain, 
the harmonic map is only invertible if the boundary parametrization fulfills specific conditions\cite{alessandrini_Invertible_2009}.
There are also variational methods to construct invertible \textquote{quasi-harmonic maps} between two simply connected domains\cite{wang_Variational_2023}.

Inspired by these works, we propose a new algorithm to find boundary conforming coordinates for arbitrarily shaped toroidal domains.
We will first construct a harmonic mapping from each poloidal cross-section of the domain to the unit disk.
Note this is the inverse of the desired mapping.
The harmonic mapping is constructed as the solution of two Laplace's equations with Dirichlet boundary conditions. In order to avoid having to discretize the domain with a mesh, we use a boundary integral method \cite{kress_Linear_1999}.
We can prove that the harmonic mapping generated this way is a diffeomorphism on the interior and thus invertible. By inverting this mapping we obtain the desired mapping from the unit disk to the cross-section of the domain.

The paper is structured as follows. In \sref{sec:method} we present the new method to construct the mapping from the 2D domain to the unit disk.
In \sref{sec:proof}, we prove the existence and regularity of the mapping, as well as that it is a diffeomorphism on the interior.
In \sref{sec:numerics} we present numerical results and finally present a discussion in \sref{sec:conclusion}.

\section{Proposed Method} \label{sec:method}
\begin{figure}[htbp!]
\vspace{2ex}\centerline{\includegraphics[width=0.75\linewidth]{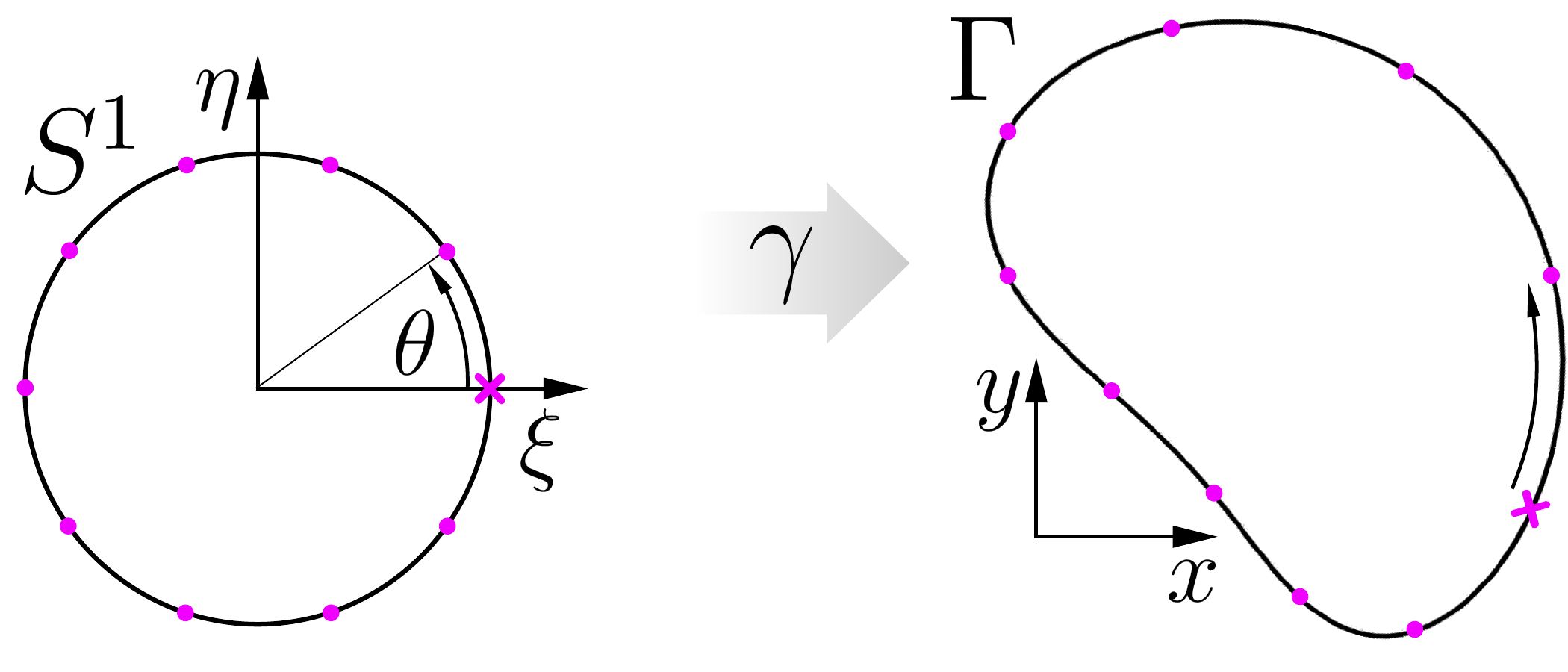}}\vspace{-2ex}
\caption{Sketch of the unit circle $\Sone$, a Jordan curve $\Gamma\subset\R^2$ and one possible parametrization $\gamma: \Sone\to\R^2$ with $\gamma\!\left(\Sone\right)=\Gamma$. Equidistant points of $\theta$ are purple, with the point corresponding to $\theta=0$ marked with a cross. \label{fig:sketch_gamma_map}}
\end{figure}

The goal of this method is to construct boundary conforming coordinates in a domain defined by a given boundary curve, 
which is assumed to be a Jordan curve $\Gamma$, i.e. a planar simple closed curve.
The domain $\Omega\subset\R^2$, delimited by the curve $\Gamma=\partial\Omega$, is thereby an open, simply connected set and we denote its closure, i.e. the domain including its boundary, with $\Omegabar = \Omega\cup\Gamma$.
The Jordan curve $\Gamma$ can be represented as the image of a function of an angle $\theta\in\left[0,2\pi\right)$, which is in practice given as a (truncated) Fourier series or spline function.
Because of the periodicity of the angle~$\theta$, it is convenient to view this function as a map defined on the unit circle $\Sone=\left\{\rme^{\rmi\theta} : \theta\in\left[0,2\pi\right)\right\}$, leading to the curve parametrization $\gamma: \Sone\to\R^2$, which is shown in \Fref{fig:sketch_gamma_map}. 
Furthermore, we require that $\gamma$ is of class $C^2$ (twice continuously differentiable), and injective, so that $\gamma$ is invertible on its image $\gamma\left(\Sone\right)=\Gamma$.

\begin{figure}[htbp!]
\vspace{2ex}\centerline{\includegraphics[width=0.75\linewidth]{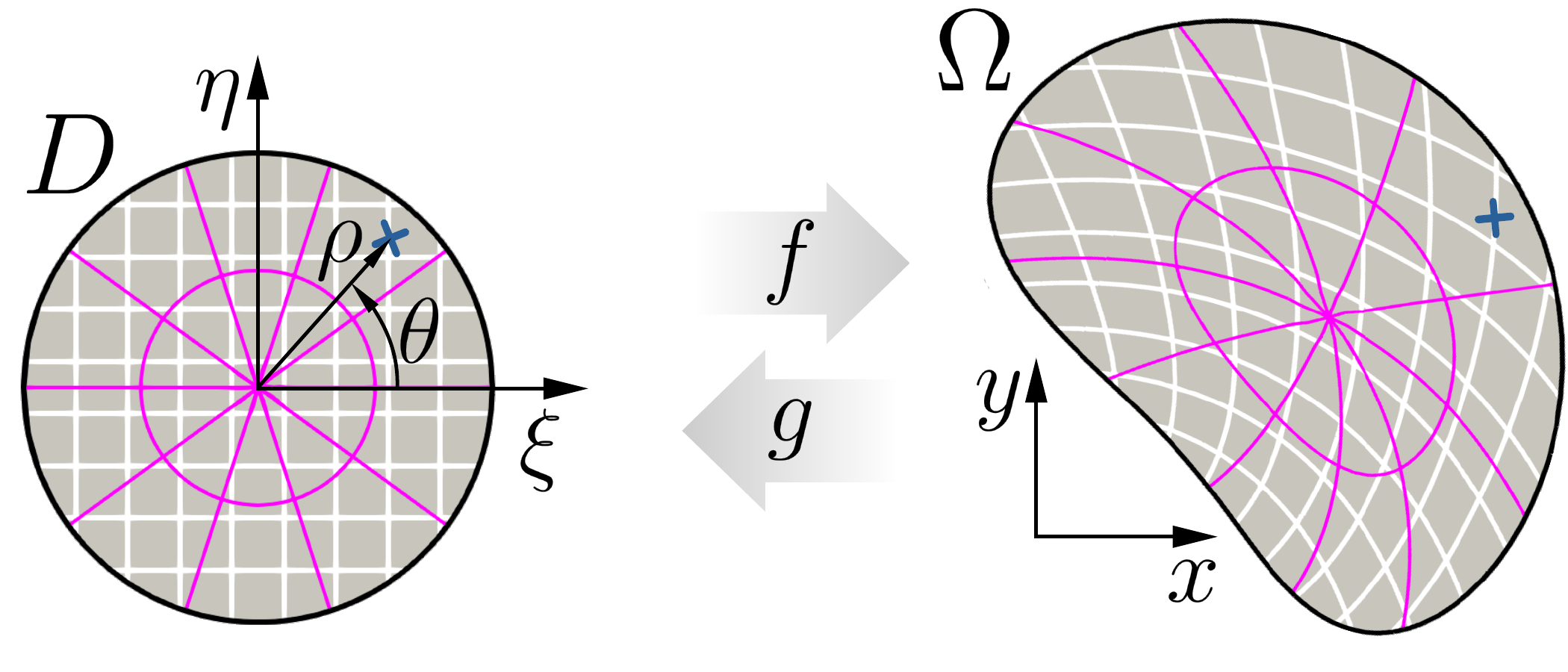}}\vspace{-2ex}
\caption{Sketch of the mapping between the (open) unit disk $\D$ and a domain $\Omega\subset\R^2$. Lines of constant $\rho,\theta$ are purple, lines of constant $\xi,\eta$ are white and the boundary is black. A blue cross marks an exemplary point in both domains, $\xi_\ell +\rmi\eta_\ell = \rho_\ell\exp(\rmi\theta_\ell)\leftrightarrow(x_\ell,y_\ell)$. \label{fig:sketch_f_g_map}}
\end{figure}

The goal of the following construction is then to find a map $f: \Dbar \to \Omegabar$ from the closed unit disk $\Dbar = \left\{z\in\C: \left|z\right|\leq 1\right\}$ to $\Omegabar$, as depicted in \Fref{fig:sketch_f_g_map}.
In order for $f$ to define valid coordinates, it must be a diffeomorphism, i.e. $f$ must be invertible and both $f$ and $f^{-1}$ must be continuously differentiable.
We will construct $f$ by inverting a harmonic map $g: \Omegabar \to \Dbar$, which is defined as the solution of two Laplace's equations with Dirichlet boundary conditions.
Solutions of the Laplace's equation on such smooth domains need to be twice continuously differentiable on the interior and continuous up to the boundary, i.e. of class $\solspace$.
As seen from the comparison of \Fref{fig:sketch_gamma_map} and \Fref{fig:sketch_f_g_map}, we additionally require that the angular parametrization of $f$, given by $z=\rhotheta$, agrees at the boundary with the given boundary parametrization $\gamma$, i.e. $f\!\left(\rme^{\rmi\theta}\right)=\gamma\!\left(\rme^{\rmi\theta}\right)$ for $\theta\in\left[0,2\pi\right)$.
For a given boundary shape, the following construction does not require any free parameters besides the parametrization of the boundary curve.

\subsection{Construction of the harmonic map}
Let $\Omega \subset \R^2$ be a simply connected domain 
bounded by a Jordan curve $\Gamma = \partial \Omega$ with $C^2$
parametrization $\gamma: \Sone \to \R^2$, $\gamma$ being invertible on its image $\gamma\left(S^1\right) = \Gamma$.
We construct the harmonic map
\begin{eqnarray} \label{eq:g_def}
    g: \Omegabar \to \Dbar,\qquad \xy \mapsto \xi\!\xy + \rmi \eta\!\xy,
\end{eqnarray}
by imposing that $\xi, \eta \in \solspace$ are solutions to the Dirichlet-Laplace problems
\numparts\begin{eqnarray}
    \Delta \xi\!\xy &= 0, \qquad& \xy\in\Omega, \label{eq:laplace-xi} \\ 
    \xi\!\xy &= \cos\theta \qquad \theta = \arg\gamma^{-1}\!\xy \qquad& \xy\in\partial\Omega, \label{eq:laplace-xi-bc}
\end{eqnarray}\endnumparts
and
\numparts\begin{eqnarray}
    \Delta \eta\!\xy &= 0, \qquad& \xy\in\Omega, \label{eq:laplace-eta} \\
    \eta\!\xy &= \sin\theta \qquad \theta = \arg\gamma^{-1}\!\xy \qquad& \xy\in\partial\Omega, \label{eq:laplace-eta-bc}
\end{eqnarray}\endnumparts
where, for $\xy \in \partial \Omega = \Gamma$, $\gamma^{-1}\!\xy = e^{i\theta} \in S^1$.
In \sref{sec:proof}, we will prove that $g|_\Omega$ is indeed a $C^2$ diffeomorphism.
Therefore the inverse $f = g^{-1}:\Dbar\to\Omegabar$ exists and is differentiable on the interior.
Transforming $\xi,\eta$ to $\rho := \sqrt{\xi^2+\eta^2}$ and $\theta := \arg\left( \xi + \rmi \eta\right)$,
where $\rho,\theta$ are polar coordinates on the unit disk, we obtain \emph{boundary conforming coordinates} for $\Omega$.
In particular, $\rho \in \left[0,1\right]$, with $\rho = 1$ on the boundary $\Gamma=\partial\Omega$, and $\theta \in \left[0,2\pi\right)$. This situation is sketched in \Fref{fig:sketch_f_g_map}.

\subsection{A simple example} \label{sec:simple-example-ellipse}
\begin{figure}[ht]
    \vspace{3ex}\centerline{\includegraphics[width=.3\linewidth]{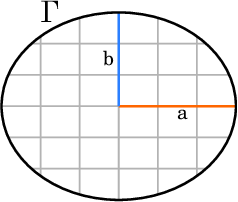}}\vspace{-2ex}
    \caption{Sketch of an elliptical domain with boundary $\Gamma$ in black, semi-major axis $a$ in orange, semi-minor axis $b$ in blue and equidistant contours of $\xi$ and $\eta$ in grey. \label{fig:sketch-ellipse}}
\end{figure}

We will now consider the simple example of an ellipse.
The boundary $\Gamma$ of the elliptical domain $\Omega$ is shown in \Fref{fig:sketch-ellipse}.
A parametrization of the boundary is given by
\begin{equation}
    x = a \cos\theta,\quad  y = b \sin\theta,
\end{equation}
where $a,b$ are the semi-major and semi-minor axes respectively and $\theta \in [0,2\pi)$ is the periodic curve parameter.
The boundary parametrization $\gamma: \Sone\to\R^2$ is then defined as
\begin{equation} \label{eq:example-ellipse-gamma}
    \gamma(\xi+\rmi\eta) = \pmatrix{
        a \cos\theta \cr
        b \sin\theta }
    = \pmatrix{a \xi \cr b \eta}, \qquad \xi+\rmi\eta = \rme^{\rmi\theta} \in \Sone.
\end{equation}

The solutions of \eref{eq:laplace-xi}-\eref{eq:laplace-eta-bc} are
\begin{equation}
    \xi\xy = \frac{x}{a}, \quad \eta\xy = \frac{y}{b}, \qquad \xy\in\Omegabar.
\end{equation}
Therefore the harmonic map $g$ is given by
\begin{equation}
    g\xy = \frac{x}{a} + \rmi \frac{y}{b}, \qquad \xy\in\Omegabar.
\end{equation}

To check whether the boundary parametrization is preserved in the harmonic map, we can compute the composition $g\circ\gamma$ for all $\xieta\in\Sone$:
\begin{equation*}
    g\circ\gamma \left(\xi+\rmi\eta\right) = g\!\left(\gamma\!\left(\xieta\right)\right) 
    = g\!\left(a \xi, b \eta\right) 
    = \frac{a \xi}{a} + \rmi \frac{b \eta}{b} = \xieta,
\end{equation*}
which is equal to the identity map, i.e. $g\circ\gamma = \Id$.

The target mapping $f:\Dbar\to\Omegabar$ is the inverse of $g$, which evaluates to
\begin{eqnarray} \label{eq:example-ellipse-f}
    f(\xieta) = \pmatrix{a \xi \cr b \eta}, \qquad& \xieta\in\Dbar, \\
    f(\rho\rme^{\rmi\theta}) = \pmatrix{a \rho\cos\theta \cr b \rho\sin\theta}, \qquad& \rho\rme^{\rmi\theta}\in\Dbar.
\end{eqnarray}
The similarity of \eref{eq:example-ellipse-gamma} and \eref{eq:example-ellipse-f}, i.e. that they only differ in the domain of definition, is specific to this example.
In general, $f$ and $\gamma$ have to agree only on the boundary, i.e.
\begin{equation}
    f\!\left(\rme^{\rmi\theta}\right) = \gamma\!\left(\rme^{\rmi\theta}\right), \qquad \theta \in [0,2\pi).
\end{equation}
It is worth noting that this mapping is not conformal, but instead it gives the intuitive linear transformation of the unit disk to the ellipse, i.e. contours of constant $\rho$ are ellipses with the same aspect ratio as the boundary.

\subsection{Implementation}
In general it is not possible to find analytic expressions for $g$.
To obtain the harmonic map $g: \Omegabar \to \Dbar$, we solve the two Laplace-Dirichlet problems \eref{eq:laplace-xi}-\eref{eq:laplace-eta-bc} using a boundary integral method \cite{kress_Linear_1999}.
We explicitly choose a boundary integral method to avoid discretizing the Laplace problem in $\Omega$.
Instead, the evaluation of the map $g$ at arbitrary points in $\Omega$ reduces to a boundary integral over the density of the corresponding double layer potential \cite[eq. 6.19]{kress_Linear_1999}.
This double layer potential can be interpreted as a continuous dipole charge distribution on the boundary $\partial\Omega$, such that the potential generated by that charge distribution is a solution of Laplace's equation with the boundary conditions \eref{eq:laplace-xi-bc} and \eref{eq:laplace-eta-bc}.
For evaluations near the boundary, a correction is needed to account for the numerical issues arising from the singularity of the double layer potential \cite{BIE2D, pyBIE2D}.
To obtain a discretized diffeomorphism $f_h \approx f = g^{-1}: \Dbar \to \Omegabar$, $\xi+\rmi\eta\mapsto \xy$ we use the following algorithm:
\begin{enumerate}
    \item Solve for the density of the double layer potential corresponding to the Dirichlet-Laplace problems~\eref{eq:laplace-xi}-\eref{eq:laplace-eta-bc}.
    This now allows evaluation of $g\!\xy = \xi\!\xy + \rmi\eta\!\xy$ at any point inside the domain, using a boundary integral.
    \item Choose a Zernike basis \cite{zernike_1934,boyd_Comparing_2011,dudt_DESC_2020} (see \ref{app:zernike}) with maximum polynomial degree~$M$ and determine the interpolation or integration points $\xi_\ell, \eta_\ell$, e.g. based on a concentric sampling pattern\cite{ramos-lopez_Optimal_2016}.
    \item Find $\left(x_\ell,y_\ell\right) \in \Omega$, such that $g\left(x_\ell,y_\ell\right) = \xi_\ell+\rmi\eta_\ell$ with the Newton-Raphson method.
    \item Solve for the coefficients of the Zernike polynomials for the map $f_h$ using the pointwise solution $\xi_\ell+\rmi\eta_\ell\mapsto (x_\ell,y_\ell)$ found in step (iii)
    \item Check the Jacobian determinant $\det f_h$ and refine the degree $M$ if necessary, e.g. if $\det f_h\left(\xi + \rmi\eta\right) < 0$ for any $\xi + \rmi\eta \in \D$ or $\min\det f_h \ll \min\det g^{-1}$,
    where $\det f_h$ and $\det g$ are the Jacobian determinants of $f_h$ and $g$ respectively.
    For the examples in \sref{sec:numerics} we evaluate the Jacobian determinant of $g$ pointwise on a regular grid in $\Omega$ and the Jacobian determinant of $g^{-1}$ or $f_h$ on a polar grid in $\D$.
\end{enumerate}

The proposed method is implemented in the \texttt{map2disc} \footnote{Available at \url{https://gitlab.mpcdf.mpg.de/gvec-group/map2disc}} python package.
To solve the Laplace problems for $\xi$ and $\eta$, we use a fast and accurate boundary integral method, as implemented in the \texttt{pyBIE2D} python package\cite{pyBIE2D}.

\section{Existence and regularity of the constructed mapping} \label{sec:proof}
In this section, we present the proof for the existence, regularity and invertibility
of $g$.

\begin{theorem}
Let $\Omega \subset \R^2$ be a simply connected domain 
bounded by a Jordan curve $\Gamma = \partial \Omega$ with $C^2$
parametrization $\gamma: \Sone \to \R^2$, with $\gamma$ invertible on its image $\gamma(S^1) = \Gamma$.
Then:
\begin{enumerate}
    \item There exists a unique map $g: \Omegabar \to \Dbar$ in $\solspace$ such that
    \begin{itemize}
        \item $g \circ \gamma = \Id$,
        \item $g$ is a harmonic function in $\Omega$,
    \end{itemize}
    \item The restriction $g|_{\Omega}$ is a $C^2$ diffeomorphism.
\end{enumerate}
\end{theorem}

\begin{remark}
This theorem differs from the Riemann Mapping Theorem in that the map $g$ is not necessarily a conformal map, 
which can be beneficial for discretization.
However, if $\gamma$ is identical to the boundary parametrization of a conformal map, then $g$ is a conformal map.
Note also that this theorem is similar to the Radó-Kneser-Choquet Theorem \cite{kneser_A41_1926} and its generalization \cite{alessandrini_Invertible_2009}, but uses a harmonic map from a simply connected domain to the unit disk and not the reverse.
\end{remark}

\begin{proof}
Let $\xi, \eta \in \solspace$ be the solutions to the Dirichlet-Laplace problems~\eref{eq:laplace-xi}-\eref{eq:laplace-eta-bc}.
From \cite[Corollary 11.13]{axler_Harmonic_2001}, it follows that the two Dirichlet problems are solvable, i.e., that $\xi, \eta$ exist and are unique.
Furthermore from \cite[Theorem 1.28]{axler_Harmonic_2001} it follows that $\xi, \eta$ are real analytic in $\Omega$.

Then by construction, $g$ as defined in~\eref{eq:g_def},
is a complex-valued harmonic function in $\Omega$ and $g \circ \gamma = \Id$, since, for any $z \in S^1$, $\theta := \arg z$,  
\begin{eqnarray}
    \nonumber
    g\circ\gamma(z) &= \xi\circ\gamma \left(z\right) + \rmi \eta\circ\gamma \left(z\right)  \\
    \nonumber
    &= \cos\theta + \rmi \sin\theta \\ 
    \nonumber
    &= \rme^{\rmi\theta} =z,
\end{eqnarray}
given the boundary conditions~\eref{eq:laplace-xi-bc} and~\eref{eq:laplace-eta-bc}.

By the inverse function theorem, a sufficient condition for $g|_\Omega$ to be a diffeomorphism is that the Jacobian matrix of $g$ is invertible, which is the case if $\nabla\xi$ and $\nabla\eta$ are linearly independent.

Let $u \in \solspace$ be any normalized linear combination of $\xi$ and $\eta$, i.e.
\begin{eqnarray}
    u := \frac{\alpha \xi + \beta \eta}{\sqrt{\alpha^2 + \beta^2}} \qquad& \left(\alpha,\beta\right) \in \R^2 \setminus \left\{\left(0,0\right)\right\},
\end{eqnarray}
where $\alpha,\beta$ are arbitrary but fixed.
This is equivalent to 
\begin{eqnarray}
    u := \xi \cos\phi + \eta \sin\phi \qquad& \phi \in [0,2\pi),
\end{eqnarray}
with arbitrary but fixed $\phi$, by defining $\phi := \arg\!\left(\alpha+\rmi \beta\right)$.

Then $u$ is a solution of the Dirichlet-Laplace problem
\begin{eqnarray}
    \Delta u\!\xy = 0 \ & \xy\in\Omega, \\
    u\!\xy = \cos\!\left(\theta - \phi\right) \qquad \theta = \arg\gamma^{-1}\!\xy \qquad& \xy\in\partial\Omega. \label{eq:u-bc}
\end{eqnarray}
From \cite[Theorem 1.1]{alessandrini_Critical_1987}, it follows that the critical points of $u$ in the interior of $\Omega$ are finite in number $K\in\mathbb{N}_0$, and denoting by $m_1,\dots,m_K$ their multiplicities, we have:
\begin{eqnarray}
    \sum_{k=1}^K m_{k} \le N - 1,
\end{eqnarray}
where $N$ is the number of relative maxima of $u|_{\Gamma}$. For the case of (\ref{eq:u-bc}), we check that $N=1$ and thus $\sum_k m_k = 0$. Therefore any linear combination of $\xi$ and $\eta$ has no critical point in $\Omega$, which implies that $\nabla \xi$ and $\nabla \eta$ are linearly independent. By the inverse function theorem, $g|_\Omega$ is then a $C^2$ diffeomorphism from $\Omega$ to $\D$.
\end{proof}

\section{Numerical experiments} \label{sec:numerics}
In order to assess the quality of a mapping $f: \Dbar \to \Omegabar$, we introduce a \emph{normalized Jacobian determinant} within each cross-section
\begin{equation} \label{eq:Jac_n}
    \Jac_n := \frac{\det f}{\max \left[\det f\right]} \leq 1,
\end{equation}
which will also be used for visualization. The Jacobian determinant of the mapping $f: \Dbar\to\Omegabar$, $\xi+\rmi\eta \mapsto \xy$ and its inverse $g:=f^{-1}$ are defined as
\begin{eqnarray} \label{eq:detf_detg}
    \det f :=\frac{\partial x}{\partial\xi}\frac{\partial y}{\partial\eta}-\frac{\partial x}{\partial\eta}\frac{\partial y}{\partial\xi}, \qquad
    \det g :=\frac{\partial\xi}{\partial x}\frac{\partial\eta}{\partial y}-\frac{\partial\eta}{\partial x}\frac{\partial\xi}{\partial y} = \left(\det f\right)^{-1}.
\end{eqnarray}
In practice, it is common to define the mapping using the polar coordinates $\rho,\theta$ for the unit disk. Then the Jacobian determinant of the boundary conforming mapping $h : \left(\rho,\theta\right) \mapsto \xy = f\!\left(\rho\rme^{\rmi\theta}\right)$ is related to $\det{f}$ by 
\begin{equation}
    \det{h} := \frac{\partial x}{\partial\rho}\frac{\partial y}{\partial\theta}-\frac{\partial x}{\partial\theta}\frac{\partial y}{\partial\rho} = \rho \det{f}.
\end{equation}
The mapping $f$ is invertible (and orientation preserving) if $\det f >0$, thus $\Jac_n>0$.
In some regards, the \textquote{best} mapping has a constant Jacobian determinant, i.e. $\Jac_n=1$ in the whole domain. 
This is motivated by affine maps of the unit disk, which have a constant Jacobian matrix and, thereby, a constant Jacobian determinant.
In particular, a large ratio between the maximum and minimum Jacobian determinant, and therefore a small $\min\Jac_n$, often leads to problems with numerical methods without mesh refinement.
On the other hand, this metric alone is not ideal as a certain \textquote{straightness} of the coordinate contours is desired \cite{tecchiolli_constructing_2024}.

\subsection{Optimized stellarator}
\begin{figure}[t]
    \centering
    \includegraphics[width=0.98\textwidth,trim=40 60 320 40,clip]{
        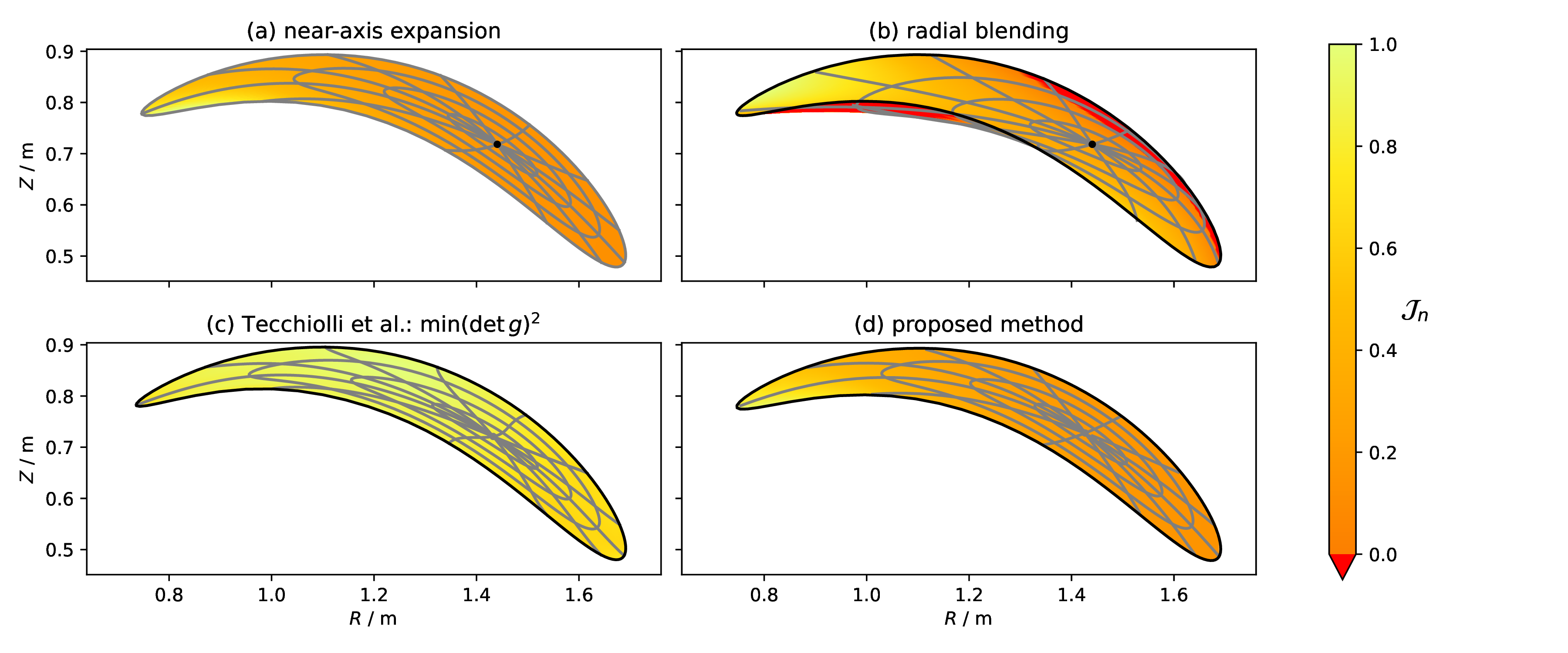    
    }
    \caption{The $\phi=\pi/8$ poloidal cross-section of the optimized two-field-period quasi-helically symmetric stellarator generated from the near-axis method\cite{landreman_mapping_2022, pyQSC_0.1.2}.
    The color shows the normalized Jacobian determinant \eref{eq:Jac_n}, with yellow depicting regions with high $\Jac_n$, orange depicting regions of low $\Jac_n$, and red depicting invalid regions with negative $\Jac_n$.
    Contours of $\rho$ and $\theta$ are grey, the prescribed boundary and/or axis is black.
    (a) is the near-axis solution.
    (b) is constructed using radial blending (see \ref{app:radial_blending}).
    (c) is constructed by minimizing the squared Jacobian determinant\cite{tecchiolli_constructing_2024} (with a straightness weighting coefficient of $\omega=0.001$)
    and (d) is constructed using the proposed method.}
    \label{fig:qh-nfp2}
\end{figure}

One example configuration where simple radial blending (see \ref{app:radial_blending}) is not sufficient is an optimized two-field-period quasi-helically symmetric stellarator configuration found using the near-axis expansion method\cite{landreman_mapping_2022}, provided as the \texttt{2022 QH nfp2} configuration of the \texttt{pyQSC}\cite{pyQSC_0.1.2} library.
As such, boundaries of this or similar configurations cannot be used with the default initialization method of VMEC and GVEC.
\Fref{fig:qh-nfp2} shows coordinate contours for the boundary conforming coordinates $\rho$ and $\theta$ on the $\phi=\pi/8$ cross-section for different initialization methods, with normalized Jacobian $\Jac_n$ shown in color.

For radial blending, overlapping coordinate contours are clearly visible, and the Jacobian determinant assumes negative values in the regions of overlapping contours.
Both the minimization of the squared Jacobian determinant\cite{tecchiolli_constructing_2024} and the proposed method show no such overlapping contours, and the Jacobian determinant is positive everywhere.
Coincidentally the central axis of both methods is close to the magnetic axis of the near-axis expansion, despite it not being specified in the construction.
For the construction by minimizing $\left(\det g\right)^2$, it was necessary to choose a \textquote{small} value of the straightness weighting factor $\omega=0.001$.
This then produces a mapping with a more uniform distribution of $\Jac_n$ than the mapping constructed with two harmonic functions, but the coordinate contours are less smooth.

\subsection{Illustrative example and convergence analysis} \label{sec:convergence}
\begin{figure}[htbp!]
    \centering
    \includegraphics[width=0.98\textwidth,trim=10 10 20 0,clip]{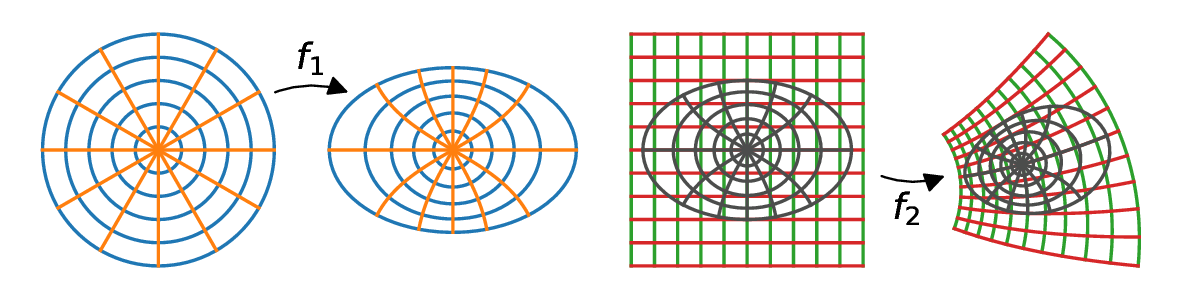}
    \caption{The two components of the conformal map $f_c=f_2 \circ f_1$. $f_1$ maps the unit disk to the ellipse and $f_2$ maps the plane to a polar domain.
    For $f_1$ the contours of $\rho$ and $\theta$ are shown in blue and orange and for $f_2$ the contours of $\xi$ and $\eta$ are shown in green and red with the $\rho$ and $\theta$ contours of the ellipse overlaid in grey.}
    \label{fig:conformal_maps}
\end{figure}

To investigate the accuracy and convergence of the proposed method, we consider a conformal map $f_c: \Dbar\to\Omegabar$ of a sheared ellipse.
The map $f_c = f_2 \circ f_1$ is shown in \Fref{fig:conformal_maps} and is composed of the conformal map of the ellipse $f_1$ (see \ref{app:confmap_ellipse}) with semi-axes $a=0.9$ and $b=0.6$ and a polar map $f_2$, representing the shear, defined as
\begin{equation}
    f_2\left(z\right) := \exp\left[\frac{z}{2} \exp\left(\frac{\rmi \pi}{12}\right)\right].
\end{equation}
As $f_2$ is a holomorphic function, it is a conformal map, and the composition $f_c$ is also a conformal map. This particular map was chosen as it features a high complexity by requiring an infinite Fourier series to represent the boundary parametrization despite having a comparatively simple shape, and it shows less symmetry than the ellipse alone.

\begin{figure}[htbp!]
    \centering
    \includegraphics[width=0.98\textwidth,trim=5 10 10 10,clip]{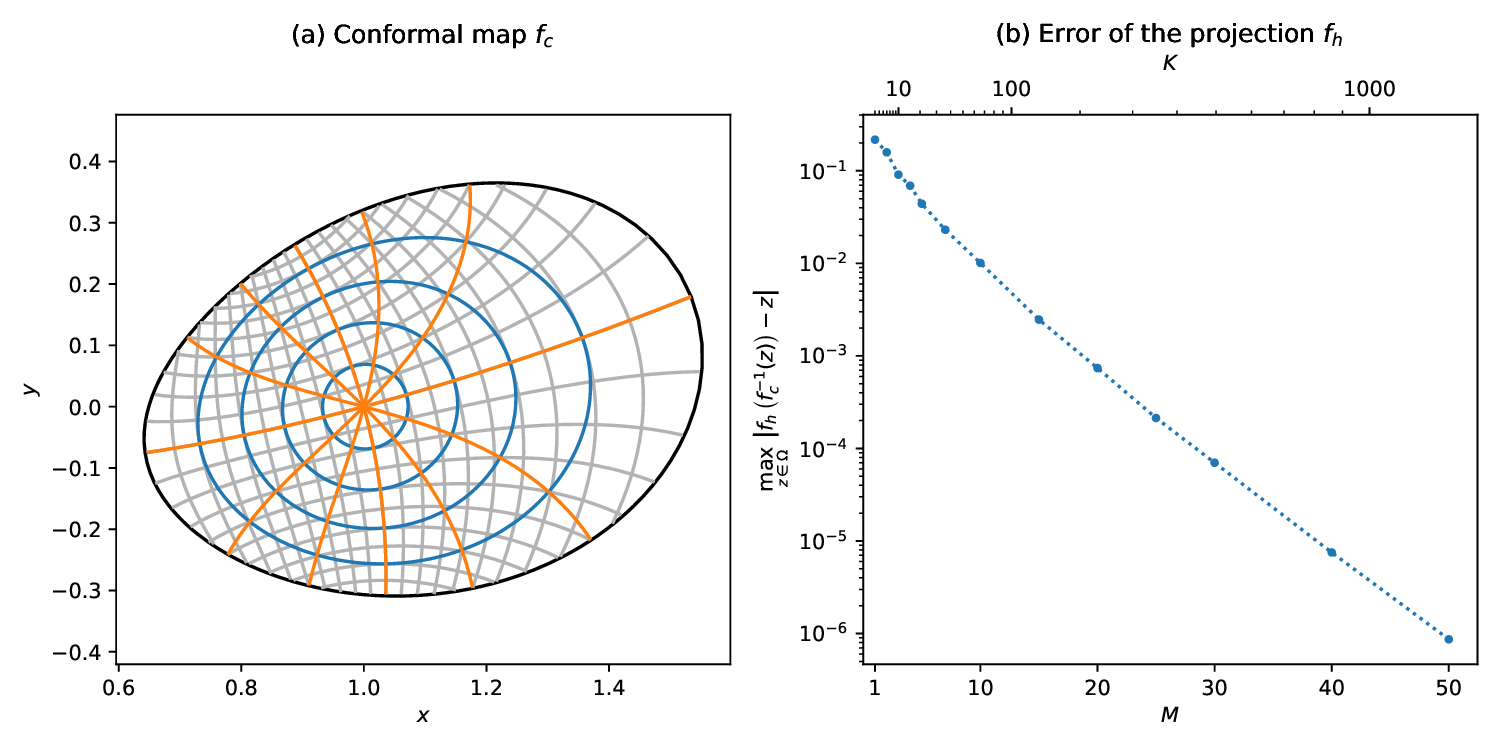}
    \caption{(a) The conformal map $f_c$ with contours of $\rho$ and $\theta$ in blue and orange respectively and contours of $\xi$ and $\eta$ in grey. (b) Maximum error of the discrete approximation $f_h \approx f_c$ by projection onto a Zernike basis with maximum degree $M$ and corresponding number of basis functions $K$.}
    \label{fig:error_projection}
\end{figure}

The implementation of the analytic conformal map $f_c$ and its inverse $f_c^{-1}$ is accurate up to machine precision, with an error of 
\begin{equation*}
    \max_{z\in\Omega}\left|f_c\left(f_c^{-1}\left(z\right)\right) - z\right| < 7.4\cdot10^{-16}.
\end{equation*}
Note that the errors were evaluated on a grid of $30076$ points in $\Omega$ (selected from a bounding box of $200\times 200$ points in $\R^2$).
The harmonic map $g$, evaluated with the boundary integral method as implemented in \texttt{pyBIE2D} with $500$ points on the boundary shows an error of 
\begin{equation*}
    \max_{z\in\Omega}\left|f_c\left(g\left(z\right)\right) - z\right| < 2.87\cdot10^{-14}.
\end{equation*}
The negligible difference between $g$ and $f_c$ shows that with our implementation, the specification of a boundary parametrization of a conformal map reproduces this conformal map.
The numerical inversion with Newton iterations (with the default tolerance of $10^{-8}$ in the $\xi + \rmi\eta$ position) shows an only marginally higher error of 
\begin{equation}
    \max_{z\in\Omega}\left|g^{-1}\left(f_c^{-1}\left(z\right)\right) - z\right| < 2.89\cdot10^{-14}.
\end{equation}
The error of the discrete approximation $f_h$ depends strongly on the number of basis functions $K$ and the maximum degree $M$, as shown in \Fref{fig:error_projection}.

\begin{figure}[htbp!]
    \centering
    \includegraphics[width=0.98\textwidth,trim=5 50 10 50,clip]{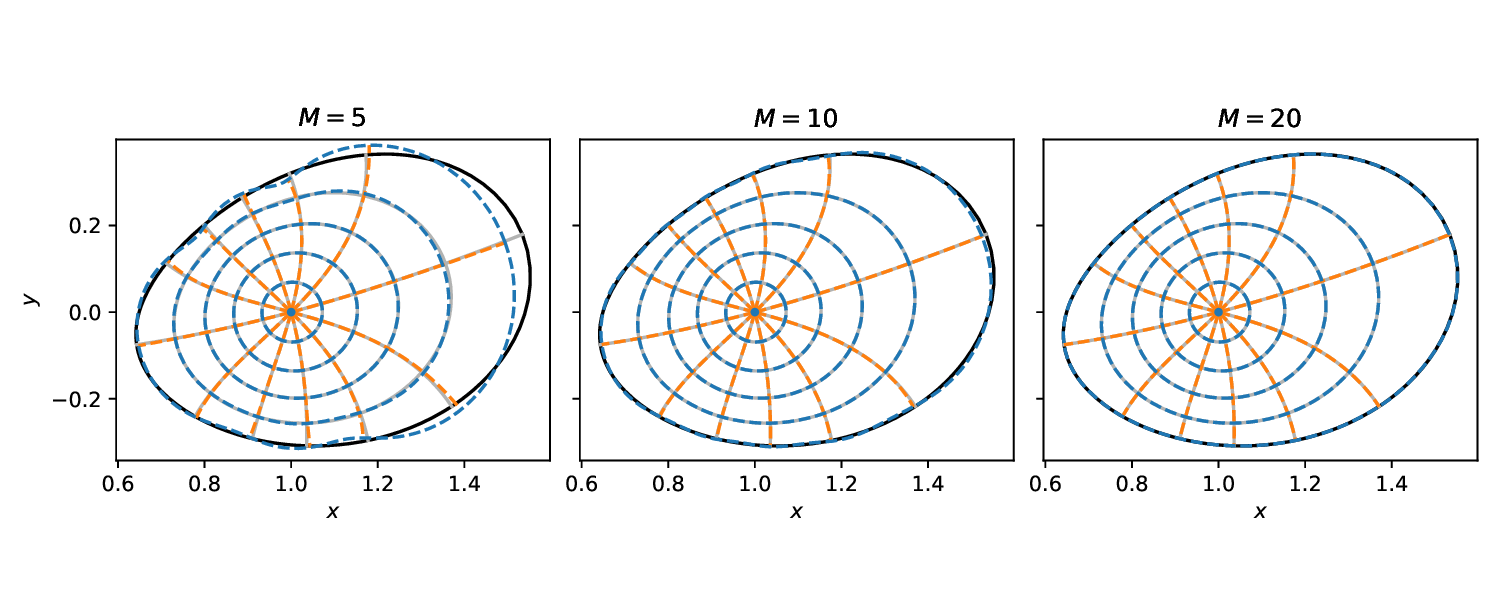}
    \caption{Discrete approximations $f_h \approx f_c$ of the conformal map of a sheared ellipse, constructed by projection onto a Zernike basis with different maximum degrees $M$. Contours of $\rho$ and $\theta$ are blue and orange for $f_h$, respectively, and grey for $f_c$. The boundary of $f_c$ is black.}
    \label{fig:conformal_discrete}
\end{figure}

\Fref{fig:error_projection} shows the conformal map $f_c$ at the left and the maximum error of the discrete approximation $f_h \approx f_c$ by projection onto a Zernike basis (see \ref{app:zernike}) with maximum degree $M$ and corresponding number of basis functions $K$ at the right.
The projection is performed using Gaussian quadrature in $\rho$ with $4(M+1)$ points and the trapezoidal rule in $\theta$ with $4(2M + 1)$ points.
In the semi-logarithmic plot, the error decreases in a manner characteristic of spectral convergence in $M$, but it only reaches values of $10^{-6}$ for $M=50$, which already requires $K=1326$ basis functions.
This low accuracy is due to the high complexity of $f_c$ and the resulting high number of Fourier modes required to accurately represent the boundary parametrization.
\Fref{fig:conformal_discrete} shows three different discrete approximations of this mapping, visually displaying the convergence seen on the right of \Fref{fig:error_projection}.
In particular, for low numbers of $M$, the mapping seems to be accurate near the axis but cannot accurately represent the boundary of the domain. This is consistent with the fact that the boundary requires many Fourier modes to be well-representable and that conformal maps tend to become circular around the axis.

\subsection{More examples}
\begin{figure}[htbp]
    \centering
    \includegraphics[width=0.98\textwidth,trim=70 40 330 30,clip]{
        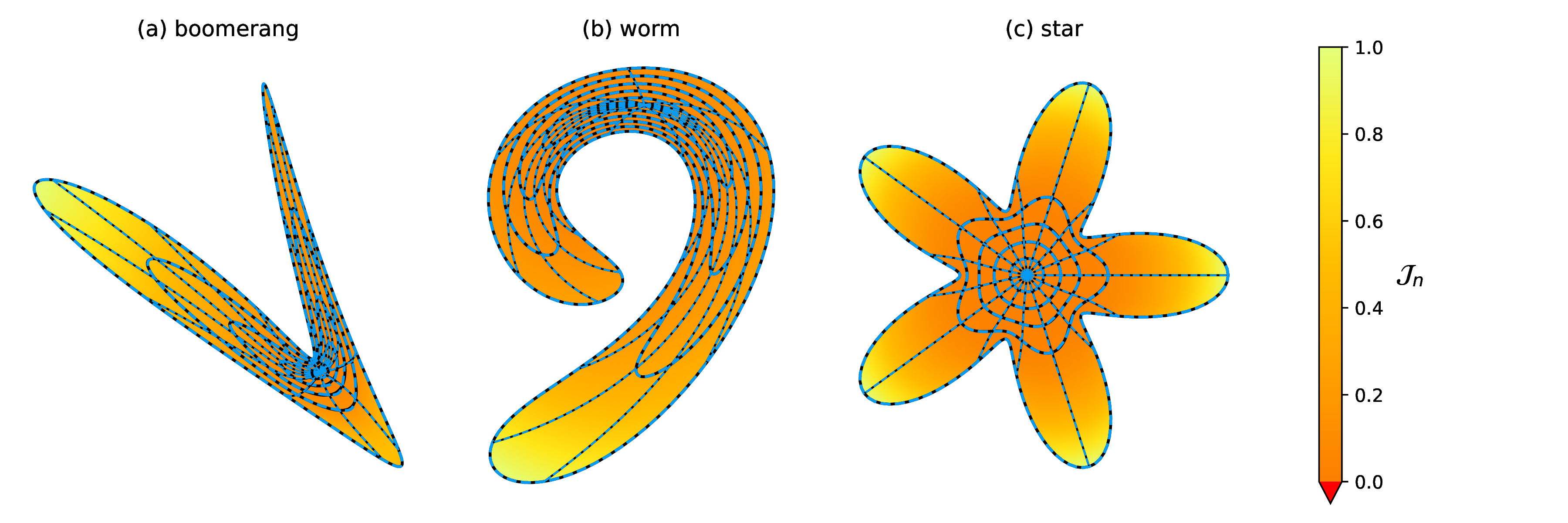
    }
    \caption{Boundary conforming mappings for three strongly shaped domains.
        Contour lines of $\rho$ and $\theta$ for the harmonic mapping $g$ are in black, and for the corresponding discrete mapping $f_h$ in blue, constructed by interpolation with a Zernike basis of maximum degree $M=15$.
        The domain is colored with the normalized Jacobian determinant \eref{eq:Jac_n} of the discrete mapping $f_h$.}
    \label{fig:examples}
\end{figure}

One of the key benefits of the proposed method is that the construction is provably invertible for arbitrarily shaped domains. 
In \Fref{fig:examples}, we show the contour lines of $\rho$ and the normalized Jacobian determinant $\Jac_n$ for the discrete approximations $f_h$ of the mapping $g^{-1}$ for three different, strongly shaped domains.
The mappings are discretized by interpolation with a Zernike basis of maximum degree $M=15$.
Even though the Jacobian determinant and, thereby, the density of contour lines can vary strongly within the domain, $\Jac_n$ is always positive, and the mapping is thus invertible.
Note that for the \textquote{star} domain, it is challenging to find a numerical conformal map \cite{trefethen_Numerical_2020}.

The mappings for the \textquote{boomerang} and \textquote{star} domains represent the boundary of the domain with machine precision, i.e. 
\begin{equation*}
    \varepsilon := \frac{1}{d} \max_{z\in\partial\Omega}\left|f_h\left(\gamma^{-1}\left(z\right)\right) - z\right| < 4\cdot10^{-15},
\end{equation*}
relative to the maximum extent $d$ of the domain $\Omega$.
The mapping for the \textquote{worm} domain, for which the boundary has been specified using B-splines, has a maximum error of the boundary representation of
\begin{equation*}
    \varepsilon < 4\cdot10^{-5}.
\end{equation*}
This error decreases with increasing maximum degree $M$, similar to the convergence of the conformal map in \Fref{fig:error_projection}.

\section{Conclusion} \label{sec:conclusion}
The presented novel method allows the construction of a mapping from an arbitrary, simply connected two-dimensional domain, bounded by a Jordan curve to the unit disk, by solving two Dirichlet-Laplace problems using a boundary integral method.
Our method therefore separates the construction of a discrete mapping into two parts:
In the first step, the harmonic mapping is generated, which can be evaluated at arbitrary points using the boundary integral and is provably invertible.
The second step is then the discrete approximation of the inverse of the harmonic map with a suitable method, e.g. by projection or interpolation of a Zernike polynomial basis.
We have proven that the harmonic mapping is a diffeomorphism on the interior by leveraging known theorems about the properties of harmonic functions. Thus the mapping is invertible and can be used to define boundary conforming coordinates.
This method does not require knowledge of the axis for the construction of the mapping.
The parameterization of the boundary and the resolution of the discrete mapping are the only free parameters in the proposed method.
The properties of the resulting map depend strongly on the parameterization of the boundary.
This has a direct influence on the perceived quality of the mapping and the resolution required to reach low approximation errors for the discrete approximation of the inverse map.
The effects of this influence can be seen in the comparison of two different parametrizations of the ellipse in \Fref{fig:sketch-ellipse} and \Fref{fig:conformal_maps}.
Hence, for future applications it could be of interest to optimize the boundary parameterization for a particular property.
Nevertheless, the proposed construction methods is guaranteed to work for any parametrization of the boundary.
If the parametrization of the boundary is chosen to be the same as the one obtained by means of a conformal map, the construction will yield exactly this conformal map.
However, we find the conformal map to be difficult to discretize using Zernike polynomials, as shown in \Fref{fig:error_projection}.
Nevertheless, the maximum error of the discretization decreases in a spectral manner, which implies that the Zernike basis is in general well suited for the discretization of such mappings.
This can also be seen in \Fref{fig:examples}, which shows three examples of strongly shaped domains where a valid mapping is constructed and discretized successfully.

One of the motivations for this work was the initialization of 3D MHD equilibrium codes, which need to find boundary conforming coordinates, given a plasma boundary.
For strongly shaped boundaries, e.g., as shown in \Fref{fig:qh-nfp2}, the simple radial blending of the boundary Fourier modes does not work.
While the proposed method works with two-dimensional domains, it can be extended to toroidal domains by applying it on a series of poloidal cross-sections.

Moreover, the proposed method can also be applied in other fields where boundary conforming coordinates for toroidal or cylinder-like domains are required.
One such application would be general-purpose field-agnostic mesh generation in strongly shaped toroidal domains, e.g. for extended-MHD codes like JOREK\cite{hoelzl_JOREK_2021,nikulsin_JOREK3D_2022}, NIMROD\cite{Sovinec_NIMROD_2004,patil_NIMSTELL_2023} or M3D-C1\cite{jardin_M3DC1_2008,zhou_M3D-C1-Stell_2021}.
The construction of a harmonic mapping and discretization using Zernike polynomials is implemented in the \texttt{map2disc} \footnote{Available at \url{https://gitlab.mpcdf.mpg.de/gvec-group/map2disc}} python package for arbitrary, simply connected 2D domains. It is intended to be extended for 3D plasma boundaries with interfaces for the VMEC and GVEC 3D MHD equilibrium solvers.

Plasma boundaries for 3D MHD configurations are typically described using Fourier series.
Therefore the discrete approximation of the constructed mapping can represent the prescribed boundary exactly, provided the maximum degree of Zernike polynomials is at least the maximum number of Fourier modes of the boundary.
Approximation errors of the discrete mapping in the interior are not as important, as long as the mapping is invertible, particularly when this mapping is used as the initialization for an MHD equilibrium solver.
Finally, some applications do not require a discrete approximation but just the evaluation of the mapping or its inverse on a number of points.
In this case, the harmonic mapping can be evaluated directly with high accuracy, and a discretization of the mapping is not necessary.

\ack
The authors would like to thank our colleagues at the Numerical Methods in Plasma Physics division at IPP Garching for the insightful discussions, in particular Alexander Hoffmann, Martin Campos Pinto, and Eric Sonnendrücker, who helped us with the formulation of the proof. 

We would also like to greatly thank David Stein from Flatiron Institute, New York, for providing the \texttt{pyBIE2D} python package \cite{pyBIE2D}, a fast, open-source python-based implementation of a 2D boundary integral solver, based on Alex Barnett's package \texttt{BIE2D} written for matlab \cite{BIE2D}. The availability and simplicity of \texttt{pyBIE2D} made this project possible in the first place and allowed us to focus our implementation effort within the \texttt{map2disc} python package on the discrete approximations of the mapping.

We would also like to thank Stuart Hudson, Joaquim Loizu and the group around SPEC for drawing our attention to the problem of initialization in 3D MHD codes and for discussions on the topic. Finally we would like to thank the reviewers for their helpful remarks.

This work has been carried out within the framework of the EUROfusion Consortium, funded by the European Union via the Euratom Research and Training Programme (Grant Agreement No 101052200 — EUROfusion). Views and opinions expressed are however those of the author(s) only and do not necessarily reflect those of the European Union or the European Commission. Neither the European Union nor the European Commission can be held responsible for them.

Robert Köberl is supported by the Helmholtz Association under the joint
research school \textquote{Munich School for Data Science - MUDS}.

\section*{Data availability statement}
The data that support the findings of this study are openly available in Zenodo at \url{https://zenodo.org/records/13318488}.
An implementation of the proposed method is available as part of the \texttt{map2disc} python package at \url{https://gitlab.mpcdf.mpg.de/gvec-group/map2disc}.

\appendix
\section{Radial blending technique} \label{app:radial_blending}
A simple technique to construct a mapping $f: \Dbar \to \R^2$ from the unit disk to a domain bounded by a simple closed curve $\Gamma$ is radial blending of the Fourier components of the boundary\cite{qu_Coordinate_2020}.
Let the boundary parametrization $\gamma: \Sone \to \R^2$, with $\gamma\!\left(S^1\right) = \Gamma$, be given by the Fourier series
\begin{equation}
    \gamma\!\left(\rme^{\rmi\theta}\right) = \sum_{m=0}^{M} \gamma^c_m \cos\!\left(m\theta\right) + \gamma^s_m \sin\!\left(m\theta\right),
\end{equation}
where $\gamma^c_m, \gamma^s_m \in \R^2$ are the Fourier coefficients and let $a \in \R^2$ be the position of the axis, i.e. $f\!\left(0\right) := a$. Then the radial blending technique constructs the mapping using polar coordinates as
\begin{equation}
    f\!\left(\rhotheta\right) = \sum_{m=0}^{M} f^c_m\!\left(\rho\right) \cos(m\theta) + f^s_m\left(\rho\right) \sin(m\theta), 
\end{equation}
where the Fourier coefficients are defined as
\begin{eqnarray}
    f^c_0\!\left(\rho\right) = a + \left(\gamma^c_0 - a\right) \rho^2 \qquad& \\
    f^s_0\!\left(\rho\right) = 0 &\\
    f^{c,s}_m\!\left(\rho\right) = \gamma^{c,s}_m \rho^m \qquad& 0 < m \leq M.
\end{eqnarray}
This mapping fulfills the boundary conditions $f\!\left(\rme^{\rmi\theta}\right) = \gamma\!\left(\rme^{\rmi\theta}\right)$ and $f\!\left(0\right) = a$, but is not necessarily invertible and can map to points outside of the boundary $\Gamma$.

\section{Zernike polynomials} \label{app:zernike}
Zernike polynomials \cite{zernike_1934,boyd_Comparing_2011,dudt_DESC_2020} are a set of orthogonal polynomials on the unit disk.
They are convenient for the discretization of smooth functions on the unit disk, as they are polynomials in the Cartesian coordinates $\xi,\eta$, but also have a simple representation in polar coordinates $\rho,\theta$.
They are defined as
\begin{eqnarray}
    Z^m_l\!\left(\rho,\theta\right) = 
    \cases{
        R^m_l\!\left(\rho\right) \cos\!\left(m\theta\right) & $m \geq 0$, \\
        R^m_l\!\left(\rho\right) \sin\!\left(-m\theta\right) & $m < 0$,
    }
    \ & m = -l, -l + 2, \dots l,
\end{eqnarray}
where the radial polynomials $R^m_l$ are defined as
\begin{equation}
    R^m_l\!\left(\rho\right) = \left(-1\right)^{(l-m)/2} \rho^m P^{m,0}_{(l-m)/2}\!\left(1-2\rho^2\right),
\end{equation}
with the Jacobi polynomials $P^{m,0}_{(l-m)/2}$.

In this work, a Zernike base of maximum degree $M$ contains the polynomials $Z^m_l$ with $l = 0, 1, \dots, M$, and $m = -l, -l + 2, \dots l$ for each $l$.
It therefore contains $K = (M+1)(M+2)/2$ basis functions.
From the definition above, it is clear that the maximum degree $M$ has a direct relation to the maximum mode number of the Fourier series in the angle $\theta$.

In this work, the interpolation points for the Zernike polynomials are chosen based on a concentric sampling pattern\cite{ramos-lopez_Optimal_2016}. They are placed on $\left\lfloor \frac{M}{2} \right\rfloor + 1$ rings, whose position is determined by Chebyshev-Lobatto points of degree $M$, defined as
\begin{equation}
    \rho_i = \cos\left(\frac{i\pi}{M}\right), \qquad 0 \leq i \leq \left\lfloor \frac{M}{2} \right\rfloor.
\end{equation}
On each ring, $2M+1-4i$ points are placed equidistantly:
\begin{equation}
    \theta_{ij} = \frac{2\pi j}{2M+1-4i}, \qquad 0 \leq j \leq 2M-4i, \qquad 0 \leq i \leq \left\lfloor \frac{M}{2} \right\rfloor.
\end{equation}

\section{Conformal map of an ellipse} \label{app:confmap_ellipse}
In this section, an analytical expression of the conformal map from the unit disk to the ellipse, as depicted in the left part of \Fref{fig:conformal_maps}, as well as the expression of its inverse, are presented. They are found in \cite{schwarz1869,schwarz1890,Szego50}.

Given an ellipse in the complex plane, centered at the origin with semi-major axis $a$ and semi-minor axis $b$, the analytical expression of the conformal map from the unit disk in the complex plane $\left|\xi+\rmi\eta\right|\leq 1$ to the ellipse reads as
\begin{equation}
    f_\textrm{ell}(\xi+\rmi\eta) = c \sin \left( \frac{\pi}{2} \frac{\textrm{F}\left(\arcsin\left((\xi+\rmi\eta)/\sqrt{k},k^2\right),k^2\right)}{\textrm{K}\left(k^2\right)} \right),
    \label{eq:confmap_ellipse_f}
\end{equation}
with the linear eccentricity $c=\sqrt{a^2-b^2}$, the Legendre incomplete elliptic integral of the first kind $\textrm{F}(\phi,m)$, the complete elliptic integral $\textrm{K}(m)=\textrm{F}(\pi/2,m)$, the elliptic parameter $m=k^2$, using the elliptic modulus computed as $k=(\vartheta_2(q)/\vartheta_3(q))^2$ from the Jacobi theta functions and the elliptic nome $q=\left(a-b\right)^2/\left(a+b\right)^2$. The $x,y$ coordinates of the ellipse in $\R^2$ are simply computed from the real and imaginary part of $f_\textrm{ell}$.

The inverse of the map  $g_\textrm{ell}=(f_\textrm{ell})^{-1}$ is expressed analytically as
\begin{equation}
    g_\textrm{ell}\!\xy = \sqrt{k} \,\mathrm{sn}\left(\frac{2}{\pi}\textrm{K}\left(k^{2}\right) \arcsin\left(\frac{x+\rmi y}{c}\right),k^{2}\right)
\label{eq:confmap_ellipse_inv}
\end{equation}
where $\textrm{sn}(u,m)$ is the Jacobi elliptic function, and $g_\textrm{ell}=\xi+\rmi \eta$ is the position on the unit disk.
An implementation of these functions using the \texttt{mpmath}\cite{mpmath_1.3.0} python package is included in the \texttt{map2disc} package.

\section*{References}
\bibliographystyle{iopart-num}
\bibliography{references}

\end{document}